\documentclass[twocolumn,amssymb,prl]{revtex4}
\usepackage{amsmath}
\usepackage[dvips]{graphicx,color}
\usepackage{dcolumn}
\usepackage{bm}
\newcounter{one}
\newcommand{\expect}[1]{\langle #1 \rangle}

\newcommand{\ket}[1]{| #1 \rangle}

\newcommand{\brakets}[3]{\langle {#1} | {#2} | {#3} \rangle}
\newcommand{\op}[1]{\hat {#1}}
\def\diff{\mathrm d}

\begin{document}


\title{Demonstration of unconditional one-way quantum computations for continuous variables}

\author
{Ryuji Ukai$^{1}$, Noriaki Iwata$^{1}$, Yuji Shimokawa$^{1}$, Seiji C. Armstrong$^{1,2}$, Alberto Politi$^{1,3}$, Jun-ichi Yoshikawa$^{1}$, Peter van Loock$^{4}$, and Akira Furusawa$^{1}$}

\affiliation{$^{1}$Department of Applied Physics and Quantum-Phase Electronics Center,
School of Engineering, The University of Tokyo,\\ 7-3-1 Hongo, Bunkyo-ku, Tokyo 113-8656, Japan\\
$^{2}$Centre for Quantum Computation and Communication Technology,
Department of Quantum Science,
Research School of Physics and Engineering,
The Australian National University,
Canberra ACT 0200 Australia\\
$^{3}$Centre for Quantum Photonics, H. H. Wills Physics Laboratory \& Department of Electrical and Electronic Engineering,\\ University of Bristol, Merchant Venturers Building, Woodland Road, Bristol, BS8 1UB, UK\\
$^{4}$Optical Quantum Information Theory Group,
Max Planck Institute for the Science of Light,
Institute of Theoretical Physics I, Universit\"{a}t Erlangen-N\"{u}rnberg,
Staudtstr.7/B2, 91058 Erlangen, Germany}

\begin{abstract}
Quantum computing promises to exploit the laws of quantum mechanics for processing information in ways fundamentally different from today's classical computers, leading to unprecedented efficiency \cite{Nielsen00,Shor94}.
One-way quantum computation, sometimes referred to as the cluster model of quantum computation,
is a very promising approach to fulfil the capabilities of quantum information processing.
The cluster model is realizable through measurements on a highly entangled
cluster state with no need for controlled unitary evolutions \cite{Nielsen06,Briegel01}.
Here we demonstrate unconditional one-way quantum computation experiments for continuous variables using a linear cluster state of four entangled optical modes.
We implement an important set of quantum operations, linear transformations, in the optical phase space through one-way computation.
Though not sufficient, these are necessary for universal quantum computation
over continuous variables, and in our scheme, in principle, any such linear transformation can be unconditionally
and deterministically applied to arbitrary single-mode quantum states.
\end{abstract}

\maketitle

The cluster model of quantum computation (QC) is a recently proposed alternative to the conventional circuit model \cite{Raussendorf01,Briegel01,Menicucci06,Zhang06,Lloyd99,Nielsen06}.
In this model, unitary operations are achieved indirectly through measurements on a highly entangled quantum state -- the cluster state.
Cluster computation is achieved through the following steps:
(1) preparation of an entangled cluster state and an input state for processing;
(2) entangling operation on these two states;
(3) measurements on most subsystems of the cluster state and feed-forward of their outcomes;
(4) occurrence and read-out of the output in the remaining unmeasured subsystems of the cluster.
Universality, i.e., realization of arbitrary unitary operations is achieved by adjusting the measurement bases,
sometimes also dependent on the results of earlier measurements
\cite{Raussendorf01,Menicucci06}.

\begin{figure*}
\centering
\includegraphics[width=16cm,clip]{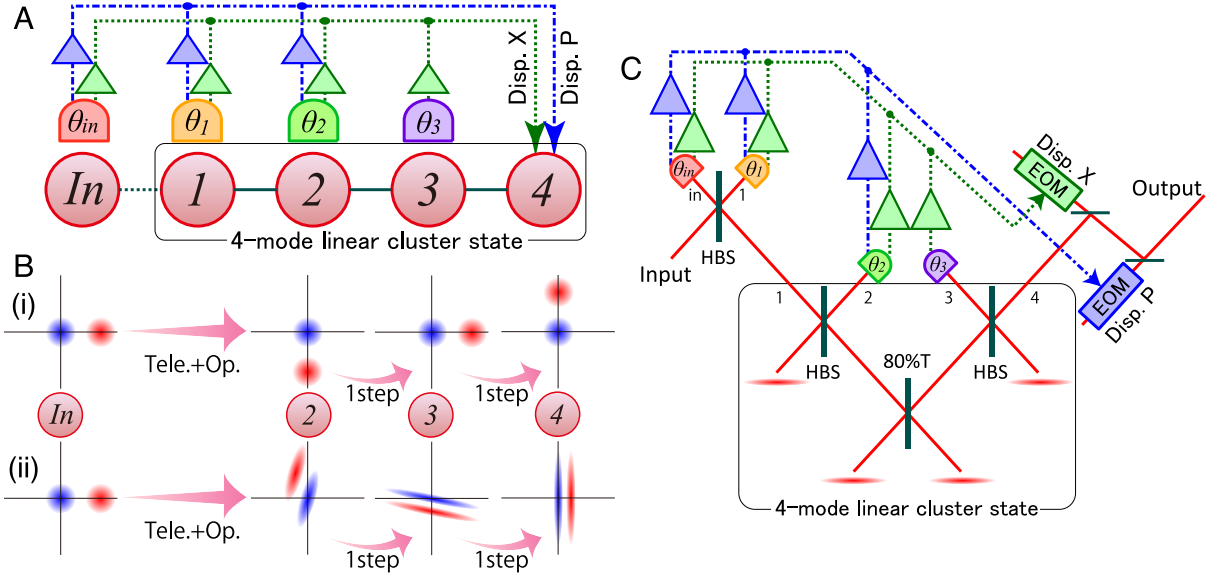}
\caption{(A) Abstract illustration and (C) experimental setup of one-mode LUBO transformations using a four-mode linear cluster state.
There is a 1-to-1 correspondence between (A) and (C).
Squeezed vacuum states are generated by subthreshold optical parametric oscillators containing
periodically poled KTiOPO$_4$ crystals as nonlinear media.
(B) phase space representations of quantum states in each step of the Fourier transformation (B-i)
and the 10dB $x$-squeezing operation (B-ii),
starting with a vacuum state input (blue) and an $x$-coherent state input (red).
Tele.: teleportation, Op: operation.
}
\label{FigSetup}
\end{figure*}

Several experiments of one-way quantum computation have been reported for discrete-variable (qubit) systems using single photons \cite{Walther05,Prevedel07,Tokunaga08,Vallone08}.
These demonstrations of one-way quantum computation work in a probabilistic way,
since the resource cluster is generated only when the photons that compose the cluster are produced and detected.
Another typical feature of the single-photon-based cluster computation experiments is that
the usual input states, $\ket{+}$, are prepared as part of the initial cluster states.
These properties would pose severe limitations when unitary gates are to be deterministically applied online
to an unknown input state which is prepared independently of the cluster state, for instance, as the output of a preceding computation.

In contrast, we report in this paper on {\it unconditional} one-way quantum computation experiments conducted
on {\it independently prepared} input states. These inputs, as well as the entangled cluster state,
are continuous-variable states.
The price to pay for this is a set of stronger requirements on universality.
Not only do we need at least one nonlinear element to achieve completely universal QC
over continuous variables \cite{Lloyd99,Bartlett}, we also have to cover all linear transformations, which,
for a single optical mode, consist of arbitrary displacement, rotation, and squeezing operations in phase space.
Our scheme represents the ultimate module for arbitrary
linear transformations of arbitrary one-mode quantum optical states.
It can be directly incorporated into a full, universal
cluster-based QC together with a nonlinear element such as measurements based on photon counting \cite{Gu09}
(for a discussion on the fidelity
when concatenating our module using finitely squeezed cluster states,
and on its scalability into a full, measurement-based QC, see
supplementary information and Refs.~\cite{Ohliger10,Browne10,NickPvL2010}).

We use a
continuous-variable
four-mode linear cluster state
as a resource \cite{Zhang06}.
An approximate version of this cluster state can be obtained deterministically by combining
four squeezed vacuum states on an 80\%-transmittance
beam splitter and two half beam splitters (HBSs) \cite{Peter07,Su07,Yukawa08C}.

Recently, it was shown that the complete set of one-mode linear unitary Bogoliubov (LUBO) transformations,
corresponding to Hamiltonians quadratic in $\op{x}$ and $\op{p}$,
can be implemented using a four-mode linear cluster state as a resource \cite{Ukai09}.
The measurements required to achieve these operations are efficient homodyne detections
with quadrature angles $\theta_i$,
which are easily controllable by adjusting the local oscillator phases in the homodyne detectors.
The total procedure then consists of the teleportation-based \cite{Vademan94,Braunstein98,Furusawa98} coupling $\op{M}_{tele}(\theta_{in},\theta_{1})$,
followed by two elementary, measurement-based, one-mode operations $\op{M}(\theta_i)$
\cite{Peter07am,Miwa09,Gu09} (see supplementary information):
\begin{align}
\ket{\psi_{out}}
=
\op{M}(\theta_3) \op{M}(\theta_2) \op{M}_{tele}(\theta_{in},\theta_{1}) \ket{\psi_{in}}.
\end{align}
Each step can be decomposed into three inner steps, a $\phi$-rotation, squeezing, and a $\varphi$-rotation in phase space:
$\op{R}(\varphi)\op{S}(r)\op{R}(\phi)$
with
$\op{R}(\theta)=e^{i\theta(\op{x}^2+\op{p}^2)}$
and
$\op{S}(r)=e^{ir(\op{x}\op{p}+\op{p}\op{x})}$
\cite{Braunstein05}.
We have
$\op{M}_{tele}(\theta_{in}$,$\theta_{1})=\op{R}(-\theta_+/2)\op{S}(r)\op{R}(-\theta_+/2)$
with $r=\log\tan(\theta_-/2)$ and $\theta_\pm=\theta_{in}\pm\theta_{1}$, while
$\op{M}(\theta_i)=\op{R}(\phi_i)\op{S}(r_i)\op{R}(\phi_i)$ with
$r_i=\log\frac{\sqrt{k_i^2+4}+k_i}{2},
\phi_i=\frac{\pi}{2}-\tan^{-1}\frac{\sqrt{k_i^2+4}-k_i}{2}$, and $k_i=1/\tan\theta_i$.



In our experiment, we demonstrate four types of LUBO transformations:
the Fourier transformation $\op{F}=\op{R}(\pi/2)$
(90$^\circ$ rotation); and
three different $x$-squeezing operations
$\op{S}(r)$ with $r=\ln 10^\frac{a}{20},a=3, 6, 10$[dB].
FIG.~\ref{FigSetup}A and FIG.~\ref{FigSetup}C show the abstract illustration and the experimental setup, respectively.
We employ the experimental techniques described in Refs.~\cite{Yukawa08C} and~\cite{Yukawa08T}
for the generation of the cluster state and the feed-forward process, respectively.

\begin{figure}
\centering
\includegraphics[width=7cm,clip]{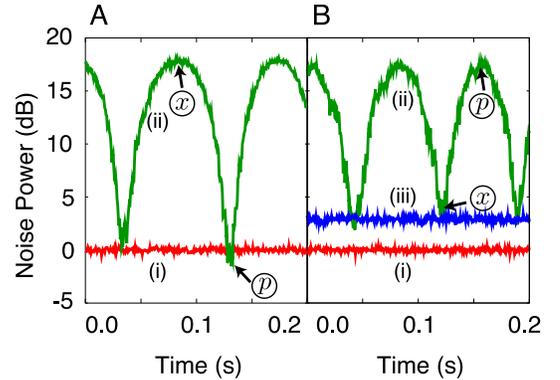}
\caption{Fourier transformation operation;
(A) Measurement results of the input state.
Trace (A-i) shows the shot noise level (SNL) and (A-ii) shows the phase scan of the input state.
(B) Measurement results of the output state.
Trace (B-i) shows the SNL,
(B-ii) shows the phase scan of the output state,
and (B-iii) shows the measurement result of the $x$ quadrature with a vacuum input.
The measurement quadrature angle is determined through the relative phase between
the signal beam and the local oscillator beam.
The measurement frequency is 1 MHz and the resolution and video bandwidths are 30 kHz and 300 Hz, respectively.
Traces (A-i), (B-i), and (B-iii) are averaged 20 times.
}
\label{FigFourier}
\end{figure}

\begin{figure*}
\centering
\includegraphics[width=17cm,clip]{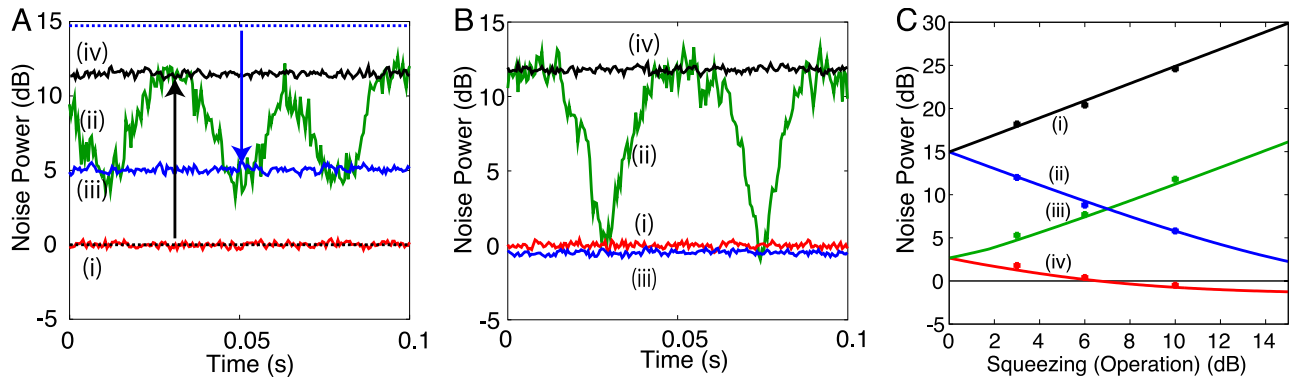}
\caption{Squeezing operations;
(A, B) 10dB $x$-squeezing operation with an $x$-coherent input(A) and a vacuum input(B).
(i) shot noise level; (ii) phase scan of the output state;
(iii) measurement of $x$; and (iv) measurement of $p$.
The measurement settings are the same as in FIG.~\ref{FigFourier}.
Traces (i), (iii), and (iv) are averaged 20 times.
(C) experimental results (dots) and theoretical calculation (solid curves) of 3dB, 6dB, and 10dB $x$-squeezing operations.
Black and blue traces correspond to a $p$ measurement with $p$-coherent input, and $x$ measurement
with $x$-coherent input, respectively;
green and red traces correspond to a $p$ measurement with vacuum input,
and $x$ measurement with vacuum input, respectively.
Each data point has an error of about $\pm$0.2dB.
}
\label{FigSqueezing}
\end{figure*}

The Fourier transformation is achieved by choosing for step (3) measurement quadrature angles
$(\theta_{in}, \theta_1,\theta_2,\theta_3)$ as $(90^\circ, 0^\circ,90^\circ,90^\circ)$, see supplementary information.

The measurement results for the Fourier transformation of a coherent state input are shown
in FIG.~\ref{FigFourier}.
As clearly shown in FIG.~\ref{FigFourier}A, the input is a coherent state with amplitude $17.7\pm 0.2$dB.
The output state is shown in FIG.~\ref{FigFourier}B.
The peak level of trace FIG.~\ref{FigFourier}B(ii) is 17.5$\pm$0.2 dB higher than the shot noise level (SNL),
which is the same level as the input within the error bar.
We acquire the peak of the input by measuring $x$,
while we obtain the peak of the output by measuring $p$,
corresponding to a 90$^\circ$ rotation in phase space.
These measurement results confirm that the Fourier transformation is applied to the input coherent state.

The quality of the operation can be quantified by using the fidelity,
defined as $F=\brakets{\Psi_{ideal}}{\op{\rho}_{out}}{\Psi_{ideal}}$.
In the specific case of our experiment, the fidelity for a coherent input state as given above is
$F=2/\sqrt{(1+4\sigma_{out}^x)(1+4\sigma_{out}^p)}$,
where $\sigma_{out}^x$ and $\sigma_{out}^p$ are the variances of the position and momentum operators
in the output state, respectively \cite{Braunstein01}.
We obtain $\sigma_{out}^x$ = 2.9$\pm$0.2 dB(FIG.~\ref{FigFourier}B(iii)), and
$\sigma_{out}^p$ = 2.8$\pm$0.2 dB (not shown) above the SNL with a vacuum input,
corresponding to a fidelity of F=0.68 $\pm$ 0.02.
This is in good agreement with the theoretical result $F=0.71$, where
an average squeezing level of $-$5.5dB is taken into account.

Another fundamental element of the LUBO transformations is squeezing.
A sequence of teleportation coupling $\op{M}_{tele}(\theta_{in},\theta_{1})$ followed by elementary one-mode one-way operations $\op{M}(\theta_i)$ is required in order to extract squeezing without rotations (see FIG. \ref{FigSetup}B(ii)).

We implemented three different squeezing operations
with three different sets of quadrature measurement angles $(\theta_{in}, \theta_1,\theta_2,\theta_3)$:
\begin{align}
\begin{array}{rc}
&(-42.5^\circ, 62.4^\circ, 63.5^\circ, 76.0^\circ), \\
&(-41.4^\circ, 72.2^\circ, 41.9^\circ, 74.4^\circ), \\
\mathrm{and} & (-47.7^\circ, 79.2^\circ, 25.9^\circ, 78.4^\circ),
\end{array}
\end{align}
resulting in 3dB, 6dB, and 10dB $x$-squeezing operations, respectively (see supplementary information).
In all these squeezing gates, the inputs are chosen to be coherent states with a nonzero amplitude in $x$ ($x$-coherent)
or in $p$ ($p$-coherent), and these amplitudes are 14.7dB$\pm$ 0.2dB.

FIG.~\ref{FigSqueezing}A shows the measurement results of the 10dB $x$-squeezing operation on the $x$-coherent state.
In this figure, the extra dotted lines are plotted for comparison, in order to show the levels of the input state:
$x$ in blue (14.7dB) and $p$ in black (SNL).
We obtain signal levels of 5.1$\pm$0.2dB and 11.5$\pm$0.2dB
above the SNL for the measurement of the $x$ and  $p$ quadratures of the output, respectively.
The level of the $x$ quadrature of the output (FIG.~\ref{FigSqueezing}A(iii))
is about 10dB lower than that of the input (the blue dotted line in FIG.~\ref{FigSqueezing}A),
while the variance of the $p$ quadrature of the output (FIG.~\ref{FigSqueezing}A(iv))
increases by about 10dB compared to that of the input (the black dotted line in FIG.~\ref{FigSqueezing}A).
These observations are consistent with a 10dB $x$-squeezing operation.
Note that the $x$ and $p$ quadratures of the output have additional noises.
These are caused by the finite squeezing of the cluster state and would vanish
in the limit of infinite cluster squeezing.

In order to show the nonclassical nature of the output state,
we also use a vacuum state as the input (FIG.~\ref{FigSqueezing}B).
The measured variance of the $x$ quadrature is $-$0.5$\pm$0.2dB,
which is below the SNL, thus confirming nonclassicality.

Finally, we demonstrate the controllability of the one-way quantum computations.
Both theoretical curves (with $-$5.5dB resources) and measured results for the three levels (3dB, 6dB, and 10dB)
of $x$-squeezing are plotted in FIG.~\ref{FigSqueezing}C.
Three kinds of input states are used here: a vacuum state; an $x$-coherent state; and a $p$-coherent state.
As can be seen in FIG.~\ref{FigSqueezing}C,
the measurement results agree well with the theoretical curves,
and all the operations are indeed controlled by the measurement bases for the four homodyne detections.

In summary, we have experimentally demonstrated one-way quantum computations with continuous variables.
All operations were perfectly controllable through appropriate choice of measurement bases for the homodyne detections.
In our scheme, arbitrary linear one-mode transformations can be applied to
arbitrary input states coming independently from the outside.
An extension to multi-mode transformations, though not demonstrated here, is also possible by similar means \cite{Ukai09}.
The accuracy of our one-way quantum computations only
depends on the squeezing levels used to create the resource cluster state.
Although in our experiment squeezing levels were sufficient to verify the nonclassical nature of the output states,
even higher levels of squeezing, as reported recently \cite{Takeno07,Mehmet09},
may lead to increased accuracies and one-way quantum computations of potentially larger size in the near future.
In order to achieve quantum operations other than linear unitary mode transformations,
nonlinear measurements besides homodyne detections would be required.
However, the demonstration of the experimental capability of
implementing an arbitrary linear single-mode transformation
through continuous-variable cluster states,
as presented here, represents a crucial step toward universal
one-way quantum computation.

\appendix

\section{Appendix A: Discrete-Variable Cluster Computations}

In the experiments reported in Refs.~\cite{Walther05,Prevedel07,Tokunaga08,Vallone08},
quantum computations are demonstrated by showing arbitrary rotations on a qubit using a discrete-variable four-qubit linear cluster state:
\begin{align}
\frac{1}{2}\sum_{\{a,b\}=\{0,1\}} (-1)^{ab} \ket{\tilde{a}}\ket{a}\ket{b}\ket{\tilde{b}},
\end{align}
where $\ket{0}$ and $\ket{1}$ are the computational basis states, while $\ket{\tilde{0}}=\ket{+}=(\ket{0}+\ket{1})/\sqrt{2}$ and
$\ket{\tilde{1}}=\ket{-}=(\ket{0}-\ket{1})/\sqrt{2}$ can be obtained with $\ket{\tilde{i}}=\op{H}\ket{i}$ and the Hadamard gate $\op{H}$.
Using this cluster state as \textit{a resource for cluster computation}
(note that, in the recent single-photon-based works,
the usual input states, $\ket{+}$, are prepared as part of the initial cluster states),
a sequence of four operations can be applied onto
an input state $\ket{\psi}_{in}$,
\begin{align}
\ket{\psi}_{out}=\op{X}_{\phi_{3}}\op{Z}_{\phi_{2}}\op{X}_{\phi_{1}}\op{Z}_{\phi_{in}}\ket{\psi}_{in},
\end{align}
where $\op{X}_{\phi_{i}}=e^{-i\phi_{i}\op{X}/2}$ and $\op{Z}_{\phi_{i}}=e^{-i\phi_{i}\op{Z}/2}$ are
$\phi_{i}$-rotations about the $X$ and $Z$ axes on the qubit's Bloch sphere
with the usual Pauli operators $\op{X}$ and $\op{Z}$, respectively.

\section{Appendix B: Continuous-Variable Cluster Computations}

In this experiment, we use a continuous-variable four-mode linear cluster state:
\begin{align}
\frac{1}{\pi}\int\!\! \diff a\,\diff b\, e^{2iab}\ket{p_1=a}\ket{x_2=a}\ket{x_3=b}\ket{p_4=b},
\end{align}
as a resource, where $\ket{x=c}$ and $\ket{p=c}=\op{F}\ket{x=c}$
(with the Fourier transformation $\op{F}=e^{\frac{i\pi}{2}(\op{x}^2+\op{p}^2)}$)
are eigenstates of the canonical conjugate position and momentum operators, respectively
$\op{x}$ and $\op{p}$, with eigenvalues $c\in \mathbb{R}$ ($\hbar=1/2$);
the subscripts label the corresponding modes.
Here, $\ket{x}$ is the computational basis for our CV system.

As is mentioned in the main text, the output state becomes
\begin{align}
\ket{\psi_{out}}
=
\op{M}(\theta_3) \op{M}(\theta_2) \op{M}_{tele}(\theta_{in},\theta_{1}) \ket{\psi_{in}} \label{EqDecomBSs}.
\end{align}
Note that $\op{M}_{tele}(\theta_{in},\theta_{1})$ cannot be decomposed into $\op{M}_1(\theta_1)\op{M}_{in}(\theta_{in})$,
because the measurements on modes {\textit{in}} and 1 are nonlocal measurements.
The operations $\op{M}_{tele}(\theta_{in}$,$\theta_{1})$ and $\op{M}(\theta_i)$ are
each elements of the one-mode LUBO transformations.

\vspace{2em}

In the following sections, we show explicit derivation of these operations and measurement quadratic angles.

\section{Appendix C: Quantum Computation using a Four-mode Linear Cluster State}

In the Heisenberg picture, a perfect four-mode linear cluster state has zero-eigenvalue correlations:
\begin{align}
\left
\{
\begin{array}{rll}
\op{p}_1^C-\op{x}_2^C &= \op{\delta}_1 & \to 0\\
\op{p}_2^C-\op{x}_1^C-\op{x}_3^C &= \op{\delta}_2 & \to 0\\
\op{p}_3^C-\op{x}_2^C-\op{x}_4^C &= \op{\delta}_3 & \to 0\\
\op{p}_4^C-\op{x}_3^C  &= \op{\delta}_4 & \to 0,
\end{array}
\right. \label{EqCluster}
\end{align}
in the limit of infinite squeezing.
The $\op{x}_j$ and $\op{p}_j$ are position and momentum operators for an optical mode $j$ with an annihilation operator
$\op{a}_j=\op{x}_j+i\op{p}_j$.
In the experiments, squeezing levels are limited, thus $\op{\delta}_i$ have non-zero variances.
An approximate four-mode linear cluster state can be generated by combining
four squeezed vacuum states on a 80\%-transmittance
beam splitter and two half beam splitters (HBSs) \cite{Peter07,Yukawa08C},
leading to the extra noise terms
\begin{align}
\left
\{
\begin{array}{rl}
\op{\delta}_1 &= \sqrt{2} e^{-r}\op{p}_1^S \\
\op{\delta}_2 &= \sqrt{\dfrac{5}{2}} e^{-r}\op{p}_3^S + \dfrac{1}{\sqrt{2}} e^{-r}\op{p}_4^S\\
\op{\delta}_3 &= \dfrac{1}{\sqrt{2}} e^{-r}\op{p}_1^S - \sqrt{\dfrac{5}{2}} e^{-r}\op{p}_2^S\\
\op{\delta}_4 &= \sqrt{2} e^{-r}\op{p}_4^S,
\end{array}
\right.
\label{EqExCluster}
\end{align}
where $\op{a}_j^S=e^{r}\op{x}_j^S+ie^{-r}\op{p}_j^S$ shows a squeezing resource for the cluster state.
We assume that each resource has a same squeezing level $r$.

Also in our scheme, an unknown input can be coupled with the four-mode linear cluster resource using a half beam splitter
utilizing the process of quantum teleportation.
The input coupling is expressed as follows:
\begin{align}
\begin{pmatrix}
\op{x}_{in}^{C'}+i\op{p}_{in}^{C'} \\
\op{x}_{1}^{C'}+i\op{p}_{1}^{C'}
\end{pmatrix}
=
\begin{pmatrix}
\frac{1}{\sqrt{2}} & -\frac{1}{\sqrt{2}} \\
\frac{1}{\sqrt{2}} & \frac{1}{\sqrt{2}}
\end{pmatrix}
\begin{pmatrix}
\op{x}_{in}+i\op{p}_{in} \\
\op{x}_{1}^C+i\op{p}_{1}^C
\end{pmatrix},
\end{align}
where $\op{a}_{in}=\op{x}_{in}+i\op{p}_{in}$ shows an annihilation operator for an input mode.

The modes $in$, 1-3 are measured simultaneously by using homodyne detections with measurement quadrature angles $\theta_j$,
and all feed-forward processes are postponed until the end of the cluster computation \cite{Peter07am,Ukai09}.
The measurement variables are
\begin{align}
\begin{pmatrix}
\op{x}_{in}^M \\
\op{x}_{1}^M \\
\op{x}_{2}^M \\
\op{x}_{3}^M
\end{pmatrix}
=
\begin{pmatrix}
\op{x}_{in}^{C'} \cos\theta_{in} + \op{p}_{in}^{C'} \sin\theta_{in}\\
\op{x}_{1}^{C'} \cos\theta_{1} + \op{p}_{1}^{C'} \sin\theta_{1}\\
\op{x}_{2}^{C} \cos\theta_{2} + \op{p}_{2}^{C} \sin\theta_{2}\\
\op{x}_{3}^{C} \cos\theta_{3} + \op{p}_{3}^{C} \sin\theta_{3}
\end{pmatrix}.
\end{align}

By using the equations above, the $x$ and $p$ quadratures of mode 4 can be expressed as
\begin{align}
\begin{pmatrix}
\op{x}_4^C \\
\op{p}_4^C
\end{pmatrix}
=
&M(k_3)M(k_2)M_{tele}(\theta_+,\theta_-)
\begin{pmatrix}
\op{x}_{in} \\
\op{p}_{in}
\end{pmatrix} \nonumber\\
&+M(k_3)M(k_2)M_{M}(\theta_+,\theta_-)
\begin{pmatrix}
\op{x}_{in}^M/\sin\theta_{in} \\
\op{x}_{1}^M/\sin\theta_{1}
\end{pmatrix}
\nonumber\\
&+M(k_3)
\begin{pmatrix}
\op{x}_{2}^M/\sin\theta_{2} \\
0
\end{pmatrix}
+\begin{pmatrix}
\op{x}_{3}^M/\sin\theta_{3} \\
0
\end{pmatrix}\label{EqInOut}\\
&
+M(k_3)M(k_2)
\begin{pmatrix}
-1 & 0 \\
0 & 1
\end{pmatrix}
\begin{pmatrix}
\op{\delta}_{1} \\
\op{\delta}_{2}
\end{pmatrix}
\nonumber\\
&+M(k_3)
\begin{pmatrix}
0 \\
\op{\delta}_{3}
\end{pmatrix}
+\begin{pmatrix}
0 \\
\op{\delta}_{4}
\end{pmatrix},\nonumber
\end{align}
where
\begin{align}
M_{tele}(\theta_+,\theta_-)=
\begin{pmatrix}
\frac{\cos\theta_- +\cos\theta_+}{\sin\theta_-} & \frac{\sin\theta_+}{\sin\theta_-} \\
\frac{-\sin\theta_+}{\sin\theta_-} & \frac{\cos\theta_+ -\cos\theta_-}{\sin\theta_-}
\end{pmatrix}, \nonumber\\
M_{M}(\theta_+,\theta_-)=
\begin{pmatrix}
-\frac{\sin\theta_- +\sin\theta_+}{\sqrt{2}\sin\theta_-} & \frac{\sin\theta_- -\sin\theta_+}{\sqrt{2}\sin\theta_-} \\
\frac{\cos\theta_- -\cos\theta_+}{\sqrt{2}\sin\theta_-} & \frac{\cos\theta_- -\cos\theta_+}{\sqrt{2}\sin\theta_-}
\end{pmatrix},  \nonumber\\
M(k)=
\begin{pmatrix}
-k & -1 \\
1 & 0
\end{pmatrix},
k_j=\frac{1}{\tan\theta_j},
\theta_\pm=\theta_{in}\pm\theta_1.
\label{EqDefs}
\end{align}
The first term on the right-hand side of Eq. \eqref{EqInOut} is the main operation controlled by measurement quadrature angles.
$M(k_i)$ and $M_{tele}(\theta_+,\theta_-)$ in Eq. \eqref{EqInOut} with Eq. \eqref{EqDefs}
correspond to $\op{M}(\theta_i)$ and $\op{M}_{tele}(\theta_{in},\theta_1)$ in Eq. \eqref{EqDecomBSs}, respectively.
The second to fourth terms correspond to back-actions of the measurements;
the quadrature operators of mode 4 are shifted depending on the measurement results $x_{j}^M$,
and these terms should be eliminated by a succeeding feed-forward process.
The remaining terms
$\left(
\begin{smallmatrix}
\op{\delta}_{x} \\
\op{\delta}_{p}
\end{smallmatrix}
\right)$
show additional components caused by imperfection of squeezing resources,
which lead to errors in cluster computations.

In our case with Eq. \eqref{EqExCluster}, the errors are
\begin{align}
\op{\delta}_{x}&=\left(\frac{1}{\sqrt{2}}-\sqrt{2}k_2k_3\right) e^{-r} \op{p}_1^S-\sqrt{\dfrac{5}{2}}e^{-r} \op{p}_2^S\nonumber\\
&\hspace{2em}+\sqrt{\dfrac{5}{2}} k_3 e^{-r} \op{p}_3^S+\frac{1}{\sqrt{2}}k_3e^{-r} \op{p}_4^S, \label{EqXPout}\\
\op{\delta}_{p}&=\sqrt{2}k_2e^{-r} \op{p}_1^S-\sqrt{\dfrac{5}{2}}e^{-r} \op{p}_3^S+
\frac{1}{\sqrt{2}}e^{-r} \op{p}_4^S,\nonumber
\end{align}

thus additional variances are
\begin{align}
\left
\{
\begin{array}{rl}
\expect{(\Delta \op{\delta}_x)^2} &= \dfrac{1}{4}e^{-2r}\left(\left(\dfrac{1}{\sqrt{2}}-\sqrt{2}k_2k_3\right)^2+\dfrac{5}{2}+3k_3^2\right)\\
\expect{(\Delta \op{\delta}_p)^2}  &= \dfrac{1}{4}e^{-2r}\left(3+2k_2^2\right).
\end{array}
\right. \label{EqAddVar}
\end{align}
Note that Eq. \eqref{EqXPout} and Eq. \eqref{EqAddVar} consist only of squeezing components derived from the squeezing resources
because all antisqueezing components are eliminated in $\op{\delta}_i$ (see Eq. \eqref{EqExCluster}).
In this case all antisqueezing components derived from the squeezing resources in the output vanish via the feed-forward process.
This leads to a considerable reduction of extra noise terms in the output.

\section{Appendix D: Derivation of Measurement Angles for the Fourier Transformation}

In order to realize the Fourier transformation
\begin{align}
\begin{pmatrix}
\op{x}_{out} \\
\op{p}_{out}
\end{pmatrix}
=
\begin{pmatrix}
0 & -1 \\
1 & 0
\end{pmatrix}
\begin{pmatrix}
\op{x}_{in} \\
\op{p}_{in}
\end{pmatrix}
, \
\end{align}
the following equation should be satisfied:
\begin{align}
M(k_3)M(k_2)M_{tele}(\theta_+,\theta_-)
=
\begin{pmatrix}
0 & -1 \\
1 & 0
\end{pmatrix}
.
\end{align}
Therefore, we get
\begin{align}
k_{in}=0,k_2=\dfrac{2}{k_1},k_3=0,
\end{align}
where $k_1$ is a free parameter which can be chosen such that
the error in the output is minimized.
Here, as can be seen from Eq. \eqref{EqAddVar}, the additional variances are
\begin{align}
\expect{(\Delta \op{\delta}_x)^2} = \dfrac{3}{4}e^{-2r},\
\expect{(\Delta \op{\delta}_p)^2} = \dfrac{1}{4}e^{-2r}\left(3+2k_2^2\right),
\end{align}
and $\expect{(\Delta \op{\delta}_p)^2}$ becomes minimal when $k_2=0$.
Thus, $(k_{in},k_1,k_2,k_3)=(0,\infty,0,0)$ is the optimal set,
which corresponds to $(\theta_{in},\theta_1,\theta_2,\theta_3)=(90^\circ,0^\circ,90^\circ,90^\circ)$.

The relation between the input and output is given by
\begin{align}
\begin{pmatrix}
\op{x}_{out} \\
\op{p}_{out}
\end{pmatrix}
&=
\begin{pmatrix}
0 & -1 \\
1 & 0
\end{pmatrix}
\begin{pmatrix}
\op{x}_{in} \\
\op{p}_{in}
\end{pmatrix}
\nonumber
\\
&\hspace{2em}+
\begin{pmatrix}
\dfrac{1}{\sqrt{2}} e^{-r} \op{p}_1^S-\sqrt{\dfrac{5}{2}}e^{-r} \op{p}_2^S \\
-\sqrt{\dfrac{5}{2}}e^{-r} \op{p}_3^S+\dfrac{1}{\sqrt{2}}e^{-r} \op{p}_4^S
\end{pmatrix},
\end{align}
and thus, the Fourier transformation is indeed achieved
and the additional variances are
\begin{align}
\expect{(\Delta \op{\delta}_x)^2} = \dfrac{3}{4}e^{-2r},\
\expect{(\Delta \op{\delta}_p)^2} = \dfrac{3}{4}e^{-2r}. \label{EqAddVer}
\end{align}

\section{Appendix E: Derivation of Measurement Angles for the $x$-squeezing Operations}
Next, we move on to $a$dB $x$-squeezing operation:
\begin{align}
\begin{pmatrix}
\op{x}_{out} \\
\op{p}_{out}
\end{pmatrix}
=
\begin{pmatrix}
10^{-\frac{a}{20}} & 0 \\
0 & 10^{\frac{a}{20}}
\end{pmatrix}
\begin{pmatrix}
\op{x}_{in} \\
\op{p}_{in}
\end{pmatrix}
. \
\end{align}
In order to achieve this operation, $k_j$ should be selected as
\begin{align}
&\hspace{5em}k_1=\frac{k_{in}}{1+2\cdot 10^{\frac{a}{20}} k_{in}},\nonumber\\
k_2&=\frac{1+10^{\frac{a}{20}} k_{in}}{k_{in}},
k_3=\frac{10^{-\frac{a}{10}}(1+10^{\frac{a}{20}} k_{in})}{k_{in}}.
\end{align}
$k_{in}$ can be chosen such that the quadrature $x$ of the output
has a minimum additional variance.
From Eq. \eqref{EqAddVar}, the optimum $k_{in}$ is independent of the levels of squeezing resources $r$,
and it is determined only by the operation level $a$.
Straightforward algebra shows that minimum variances occur when $k_{in}$ is
\begin{align}
\left
\{
\begin{array}{rlc}
k_{in} &= -10^{-\frac{a}{20}} & \left(10^{\frac{a}{10}}\leq\frac{3}{2}\right)\\
k_{in\pm} \!\!\!\!&= \dfrac{2\left(-2\cdot 10^{\frac{a}{20}}\pm\sqrt{-3+2\cdot 10^{\frac{a}{10}}}\right)}{3+2\cdot 10^{\frac{a}{10}}} & \left(10^{\frac{a}{10}}>\frac{3}{2}\right).
\end{array}
\right.
\end{align}
Both $k_{in+}$ and $k_{in-}$ give us the same variance and $k_-$ is selected for our experiment.
Therefore, the angles $(\theta_{in},\theta_1,\theta_2,\theta_3)$:
\begin{align}
\begin{array}{rc}
&(-42.5^\circ, 62.4^\circ, 63.5^\circ, 76.0^\circ), \\
&(-41.4^\circ, 72.2^\circ, 41.9^\circ, 74.4^\circ), \\
\mathrm{and} & (-47.7^\circ, 79.2^\circ, 25.9^\circ, 78.4^\circ),
\end{array}
\end{align}
should be selected for 3dB, 6dB, and 10dB $x$-squeezing operations, respectively.

\section{Appendix F: Remarks on Fidelity and Scalability}\label{SecFourierScalability}

{\it Fidelity}-- when our module for arbitrary linear one-mode transformations
is concatenated as needed for larger quantum computations,
the noise originating from the use of realistic, finitely
squeezed cluster states will lead to an accumulation of
errors and a decreasing fidelity for the output state.
This effect is inevitable and will occur even when all the remaining
operations including the homodyne measurements are performed with
100\% efficiency.

The quality of a cluster-based operation starting with an initial pure state
can be quantified using the fidelity defined as
\begin{align}
F=\brakets{\Psi_{ideal}}{\op{\rho}_{out}}{\Psi_{ideal}},
\end{align}
where $\ket{\Psi_{ideal}}$ and $\op{\rho}_{out}$ are the ideal pure output state
and the density matrix of the experimental output state, respectively.
We shall consider the measurement-based application of Fourier transformations
starting with an arbitrary pure Gaussian state, and we obtain
\begin{align}
F=\dfrac{2}{\sqrt{(1+4\sigma_{out}^x)(1+4\sigma_{out}^p)}},\label{EqFidelity}
\end{align}
where $\sigma_{out}^x$ and $\sigma_{out}^p$ are
the variances of the position and momentum operators in the output state, respectively.
Since the excess variances for a Fourier transformation, realized through
four elementary measurement-based steps, are given by Eq. \eqref{EqAddVer},
the fidelity becomes
\begin{align}
F=\dfrac{1}{1+\frac{3}{2}e^{-2r}}.
\end{align}

We can now easily extend this discussion to the fidelity for an $n$-step
teleportation (or, more specifically, a Fourier transformation, assuming $n$ is even). In this case,
we have an input coupling through teleportation (two steps) followed by an
$(n-2)$-step one-mode one-way gate with the fidelity,
\begin{align}
F=
\left
\{
\begin{array}{ll}
\dfrac{1}{1+\frac{k+1}{2}e^{-2r}}, & n=2k,\\
\dfrac{1}{\sqrt{1+\frac{k+1}{2}e^{-2r}}\sqrt{1+\frac{k+2}{2}e^{-2r}}}, & n=2k+1.
\end{array}
\right.
\end{align}
Note that a cluster-based $n$-step quantum teleportation roughly corresponds to an
$\frac{n}{4}$-step sequential quantum teleportation, because the fidelity
for an $m$-step sequential quantum teleportation is $F=\frac{1}{1+me^{-2r}}$.
The excess noise of one-way QC through a linear chain is now roughly given
by $ne^{-2r}$.
Therefore, for a larger computation with an increasing $n$ value, but an unchanged output fidelity,
the required level of resource squeezing ($e^{+2r}$) is roughly proportional to $n$.
In other words, using the variance $~e^{+2r}$ as a figure of merit
(the so-called {\it accuracy of the cluster state} \cite{Gu09}),
there is a linear (and not an exponential) dependence of the required
accuracy on the length of the computation.
Further, any desired accuracy can be achieved for a cluster state of arbitrary
size with the same squeezing levels, provided the connectivity (the maximum number of nearest neighbors
for any mode of the cluster) is constant like in the present example of a linear chain \cite{Gu09}.
Similarly, the entanglement between any mode and the remaining modes of the cluster state
for a given squeezing level only depends on the number of links for that single mode and is independent
of the size of the cluster \cite{NickPvL2010}.
However, note that the noise accumulated in a computation and hence the {\it accuracy of the computation},
for a given accuracy of the cluster resource, does depend on the length of the computation
and hence on the size of the cluster, as described above.
So there is a distinction between the accuracy of the cluster and the accuracy of the computation;
the former can be independent of the size of the cluster, whereas the latter, of course, is not.

{\it Scalability}-- we may consider the result of cascaded quantum teleportations through a linear
cluster chain with a certain finite error specified by a lower bound $F_0$ on the fidelity.
In this case, we obtain for the number of possible steps,
\begin{align}
n= 4e^{+2r}\left(\frac{1}{F_0}-1\right)-1,
\end{align}
omitting the parity of $n$ here for simplicity.
Now considering the ``classical limit'' of teleportation, $F_0=\frac{1}{2}$ \cite{Hammerer},
as a benchmark and a squeezing level of about $-5.5$dB like in our experiment,
$n=13$ steps of elementary teleportations would be possible.
A recently reported squeezing value of $-12.7$dB \cite{Eberle10}
would enable us to cascade the teleportations up to 73 times.
This alone shows already that in a weak sense (see below), our scheme is scalable and can be extended
to a higher number of quantum operations (than just the present four operations)
with the same experimental parameters and squeezing resources.

Finally, we shall briefly comment on the full scalability of the present experimental scheme,
and measurement-based (MB) QC over continuous variables in general.
An undeniable fact is that there is no proof for scalability of continuous-variable QC
in the presence of errors in a strict sense.
Strict here means that an analogous result to the so-called threshold theorems
in the discrete-variable regime is still lacking for continuous variables.
While it has been shown that arbitrarily long qubit computations (in a circuit model)
can be achieved to any desired accuracy provided the elementary components of the computation
are less faulty than a certain threshold value \cite{Nielsen00},
such a result does not exist for continuous variables.
Mapping the circuit-based thresholds to measurement-based thresholds is also possible
in the discrete-variable regime for certain abstract error models \cite{NielsenDawson,AliferisLeung}.
However, these error models do not directly apply to
those realistic, experimental situations in which cluster states are prepared
highly probabilistically through parametric down conversion (PDC) by means of linear optical
elements \cite{Walther05,Prevedel07,Tokunaga08,Vallone08}.

In the continuous-variable case, even if there was a circuit-based threshold theorem,
the transition from the circuit to the MB model appears to be fundamentally
different compared to the discrete-variable case. The reason is that the continuous-variable
Gaussian cluster states are intrinsically noisy due to their finite squeezings,
always resulting in squeezing-induced errors in a MBQC (see above). In a circuit computation,
such errors would not occur,
except for a possible encoding step into, for instance, an approximate position eigenstate.

There are now a couple of recent investigations into the effects of finite squeezing on
the scalability of MBQC with Gaussian cluster states.
In Ref.~\cite{Ohliger10}, it is argued that the accumulation of squeezing-induced
errors prevents scalability in the strict sense of arbitrarily long computations,
as long as no extra tools such as quantum error correction codes are incorporated 
from the beginning.
More precisely, when nonclassical correlations in form of an entangled state are to be transmitted
through a linear continuous-variable cluster chain, the entanglement will decay exponentially
with the length of the chain, similar to the effect of cascaded entanglement swapping
with two-mode squeezed states \cite{Peter02}. The results of Ref.~\cite{Ohliger10}, however,
are even more general than the simplest case of homodyne-based swapping along a linear chain,
including as well non-Gaussian measurements.
Contrary to these negative observations concerning scalability,
it was also shown recently that an important necessary requirement
for a cluster state to be an {\it efficient universal} resource \cite{vandenNeest}
can be indeed satisfied by the Gaussian, finitely squeezed cluster states \cite{Browne10}.
Similar to this more optimistic view, one should also note that, even though
the entanglement in a linear Gaussian cluster chain does exponentially decay,
there is no need to use an initial resource squeezing $r$ that grows exponentially
with the total number of measurement-based computation steps. In fact, the exponential
decay rate itself depends on $r$, establishing a quantitative link between
the necessary initial squeezing resources and the maximal number of operations
$n_{max}$ that still result in an effective output squeezing $r_{out}$ 
above an arbitrary constant bound $c$, $r_{out}>c$, namely \cite{UkaiNew}
\begin{align}
n_{max} \simeq -\frac{1}{2}\, e^{+2r}\, \ln\tanh c\,.
\end{align}
Therefore, in particular, we have $2r \simeq \ln(n_{max}) + const.$,
corresponding to a logarithmic increase of the initial squeezing 
with the maximal number of operations that still satisfies
a certain accuracy threshold. Note that this statement can be
equivalently made in terms of an entanglement measure
such as the so-called logarithmic negativity  
which for a pure two-mode squeezed state is proportional to the
squeezing parameter. In this case, we obtain for the initial 
resource entanglement $E$ the scaling property
$E \simeq \ln(n_{max}) + const.$ to guarantee that
the output entanglement satisfies 
$E_{out}>c$ for some accuracy bound $c$.
The bottom line of our discussion here is that
the required input squeezing and entanglement, to make sure that
the output entanglement along an arbitrarily long cluster chain
remains at least as large as some fixed bound, scales 
logarithmically with the length of the chain.

Finally, compared to the existing theoretical results on discrete-variable fault-tolerant QC
and the published experiments on single-photon-based qubit MBQC,
even the results of Ref.~\cite{Ohliger10} do not
rule out the possibility of a fault-tolerant version of MBQC over continuous variables
(see also the final discussion in Ref.~\cite{Ohliger10}).
However, in order to deal with finite-squeezing errors from the start, there is no
known error correction scheme shown to be capable of suppressing such errors
(at least for a logical continuous-variable state);
it is only known that such an encoded scheme must be based upon some nonlinear,
non-Gaussian element \cite{Niset09}, similar to the nonlinear measurement which would be needed
to achieve universal operations through the cluster state.



\textbf{Acknowledgements}
This work was partly supported by SCF, GIA, G-COE, and PFN commissioned by the MEXT of Japan,
the Research Foundation for Opt-Science and Technology, and SCOPE program of the MIC of Japan.
P. v. L. acknowledges support from the Emmy Noether programme of the DFG in Germany.
S. A. acknowledges financial support from the IARU office at the Australian National University.

\end{document}